\documentclass[preprint,preprintnumbers,nofootinbib,aps,twocolumn,10pt]{revtex4-1}

\usepackage{amsmath,amssymb,bm,epsfig}
\usepackage{color}
\usepackage{natbib}
\usepackage{hyperref} 
\usepackage{ulem}
\usepackage{graphicx}
\RequirePackage{lineno}

\usepackage{xcolor,colortbl}
\usepackage{pifont}

\def \beq{\begin{equation}}
\def \eeq{\end{equation}}
\def \beqa{\begin{eqnarray}}
\def \eeqa{\end{eqnarray}}
\def \la{\langle}
\def \ra{\rangle}
\def \l{\left(}
\def \r{\right)}

\usepackage{xspace}

\begin{document}

\title{Separation of flow from chiral magnetic effect in U+U collisions using spectator 
asymmetry}
 
\author{Sandeep \surname{Chatterjee}}
\email{sandeepc@vecc.gov.in}
\author{Prithwish \surname{Tribedy}}
\email{ptribedy@vecc.gov.in}
\affiliation{Variable Energy Cyclotron Centre, 1/AF Bidhannagar, 
Kolkata, 700064, India}  

\begin{abstract}
We demonstrate that the prolate shape of the Uranium nucleus generates anti-correlation 
between spectator asymmetry and initial state ellipticity of the collision zone, providing 
a way to constrain the initial event shape in U+U collisions. As an application, we show 
that this can be used to separate the background contribution due to flow from the signals 
of chiral magnetic effect.
\end{abstract}
\maketitle

The hot and dense QCD plasma produced in heavy ion collisions (HIC) can give rise to 
metastable vacuum with non trivial topology of the gauge field configurations like 
sphalerons and instantons. These give rise to $P$ and $CP$ odd interactions between 
quark and gluon fields that change the quark chirality~\cite{Kharzeev:1998kz,Kharzeev:2004ey}. 
In the early stages of HICs, strong magnetic fields$\sim m_{\pi}^2$ are expected to be 
produced~\cite{Kharzeev:2007jp,Skokov:2009qp,Bzdak:2011yy}. This unique combination of 
local $P$ and $CP$ odd domains amidst strong magnetic field in HIC experiments is expected 
to give rise to many interesting phenomena like the chiral magnetic effect 
(CME)~\cite{Fukushima:2008xe}. The separation of charged hadrons along the direction of 
the magnetic field has been suggested as a possible signal of CME in which the like sign
charges are expected to be emitted in the same direction~\cite{Kharzeev:2004ey}. However, 
such angular correlations can also arise due to non-CME effects like elliptic flow, resonance 
decays, momentum conservation kinematics etc~\cite{Pratt:2010zn}. In order to reduce the 
non-flow effects, the following charged particle correlator was proposed in 
Ref.~\cite{Voloshin:2004vk}
\beq
\gamma^{ab}=\la cos\l \phi^a + \phi^b -2\psi_{RP}\r\ra.
\label{eq.cosab}
\eeq
Here $\phi$ is the azimuthal angle of the particle and $a$, $b$ $=\pm$, is its charge state. 
$\psi_{RP}$ is the reaction plane angle. The non-flow effects that are random with respect 
to the reaction plane are 
eliminated by the design of this observable. It is however difficult to reduce the background from 
elliptic flow. The elliptic flow $v_2$ is largely characterised by the initial shape of the collision 
zone and the strength of CME signal depends on the number of spectators. It turns out that the 
attempts to reduce flow, say by going towards central events, also reduces the magnetic field and 
therefore the signal of CME due to decrease in the number of spectators. Therefore, although $\gamma^{ab}$ 
has been measured in Au+Au, Cu+Cu and Pb+Pb collisions~\cite{Adamczyk:2014mzf,Abelev:2012pa,Abelev:2009ad,
Abelev:2009ac}, the final verdict on CME is still not out. This is mainly because it is 
not possible to separate the background flow effects from $\gamma^{ab}$ unambiguously by 
conventional approaches~\cite{Bzdak:2009fc}.

There have been suggestions on disentangling the CME from flow~\cite{Voloshin:2010ut,Bzdak:2011np}. 
In Au+Au collisions, it has been suggested that within a narrow centrality bin large 
fluctuation of the initial state ellipticity produces a broad event by event 
distribution of $v_2$ while CME is expected to be nearly constant because of similar number of 
spectators~\cite{Bzdak:2011np}, thereby disentangling the two effects.  

The other approach is to study the collisions of deformed nucleus such as Uranium~\cite{Voloshin:2010ut}. 
It has been pointed out that in full overlap U+U collisions, the magnetic field in the overlap 
zone nearly vanishes, although a large anisotropy is generated from certain configurations of the prolate 
shape of the Uranium nucleus. This allows one to turn off CME while having substantial $v_2$ in such 
collisions~\cite{Voloshin:2010ut}. 

  \begin{figure*}[htb]
  \begin{center}
    \scalebox{1}{
   \includegraphics[width=0.8\textwidth]{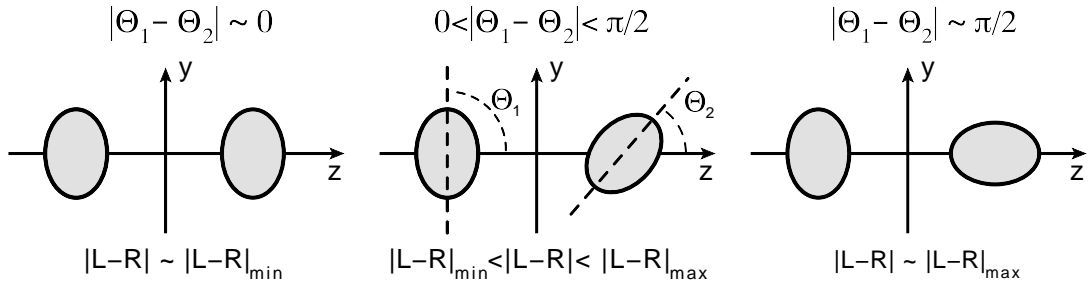}
   }
  \caption{Geometries of U+U collisions that generate higher spectator neutron asymmetry $|L-R|$ due to 
  increase in the angle between the major axes of two nuclei at a fixed impact parameter. The collision 
  direction is chosen along the z-axis. 
  }
  \label{fig.scheme}
  \end{center}
  \end{figure*}

In this work we propose a new method to systematically reduce the initial anisotropy that contributes to 
$v_2$ in a given sample of events without reducing the effect of magnetic field that generates the 
signals of CME in U+U collisions. For further discussions we introduce a quantity, the spectator nucleon 
asymmetry $|L-R|$ which is defined as the absolute difference between the left (L) and right (R) going 
nucleons that did not participate in the collision. In case of the collisions of non-deformed nuclei like 
Pb, the spherical shape of the individual nucleus ensures that $|L-R|$ receives no contribution from the 
collision geometry. The non-zero values of $|L-R|$ in such cases can arise only due to quantum fluctuations 
of the nucleon positions in the colliding nuclei. Here we will not focus on such initial state fluctuations. 
In the case of U+U collisions, $|L-R|$ receives a dominant contribution from the geometric fluctuations. 
We argue and demonstrate through Monte Carlo Glauber (MCG) simulations that $|L-R|$ is an important tuning 
parameter to constrain the initial state geometry in U+U collisions that can be useful for several purposes.

Deformed nuclear collisions are characterized by four angles representing the orientations of the major 
axes of the colliding nuclei in addition to the impact parameter b ~\cite{Voloshin:2010ut,Masui:2009qk,
Hirano:2010jg,Haque:2011aa}. They include the two polar angles $\Theta_{1}$ and $\Theta_{2}$ relative to 
the collision direction (z-axis) as shown in Fig.~\ref{fig.scheme} and two azimuthal angles $\Phi_{1}$ 
and $\Phi_{2}$ in the plane transverse to the collision direction, the collision plane. There are 
collision configurations where both the nuclei have similar projected geometry on the collision plane 
as in the case of collisions of non-deformed nuclei. Such configurations are not expected to generate large 
spectator asymmetry. On the other hand, configurations in which one nucleus largely engulfs the other 
can generate higher values of $|L-R|$. In such configurations, there is a large difference in 
the projected geometry of the two nuclei on the collision plane. In Fig.~\ref{fig.scheme} we 
demonstrate how $|L-R|$ changes with collision configurations. We start with a special configuration 
where both nuclei have the same values of $\Theta$ at a fixed $b$. In this configuration 
$|L-R|$ is generated solely due to nucleon scale quantum fluctuations. Thus, $|L-R|$ has a minimal 
value of say $|L-R|_{\text{min}}$. As the difference in the projected geometry of the two 
nuclei grows with increasing polar angle difference ($|\Theta_{1}-\Theta_{2}|$), 
we expect larger values of $|L-R|$. Finally, $|L-R|$ reaches a maximum value $|L-R|_{\text{max}}$ when the 
major axis of one nucleus aligns along the collision direction while the other one is perpendicular to it. 
We call such configurations as Body-tip (or Side-tip) collisions. Due to prolate shape of Uranium nucleus, 
it is evident that such configurations will lead to smaller anisotropy of the overlap zone. We will 
demonstrate that in a given multiplicity bin, increasing $|L-R|$ leads to systematic decrease in ellipticity. 
It is interesting to note that in case of significant oblate deformation, an opposite behavior is expected- 
higher values of $|L-R|$ will lead to larger ellipticity. 
  \begin{figure}[htb]
    \scalebox{1}{
    \includegraphics[width=0.48\textwidth]{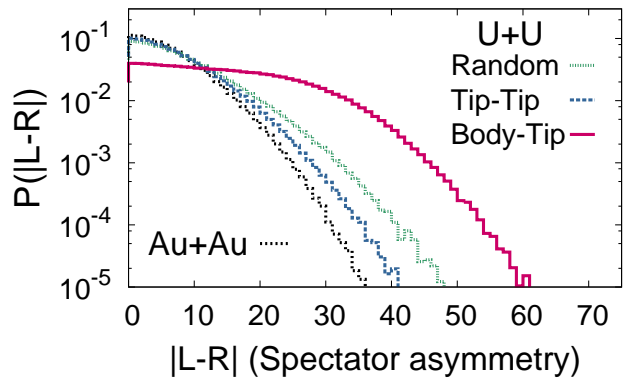}
    }
  \caption{(Color online) Probability distribution of  
  spectator neutron asymmetry parameter $|L-R|$ for Au+Au and different configurations of U+U 
  collisions.
}
  \label{fig.norm_dist}
  \end{figure}
In HIC experiments, spectator neutrons from the left and right moving nuclei are measured by two Zero Degree 
Calorimeters (ZDC) while the spectator protons are never measured~\cite{gang_wang}. Therefore a quantity of 
experimental interest would be the asymmetry of spectator neutrons. Such a quantity can be easily measured 
in the experiment by measuring the energy depositions in the two ZDCs due to incident neutrons. In the 
remaining discussion, we will refer to $|L-R|$ as the spectator neutron asymmetry.

Minimum bias events (all values of impact parameter) for U+U and Au+Au collisions at 193 GeV and 
200 GeV respectively are generated using a two-component Monte-Carlo Glauber (MCG) model. The details of 
the Glauber model and the parameters of the Wood-Saxon distributions for the nuclei used here are the same 
as used in Ref.~\cite{Schenke:2014tga}. The free parameters of the MCG model that relates collision geometry 
to multiplicity have been tuned to reproduce the min-bias multiplicity distribution predicted by the IP-Glasma 
model as described in Ref.~\cite{Schenke:2014tga}. In Fig.~\ref{fig.norm_dist}, we plot the probability 
distributions of $|L-R|$ for Au+Au collisions and three different collision configurations of U+U. The Tip-Tip 
configuration corresponds to $\Theta_{1}=\Theta_{2}=0$ and $\Phi_1=\Phi_2=0$ and the Body-Tip configuration 
corresponds to $\Theta_{1}=\pi/2, \Theta_{2}=0$ and $\Phi_1=\Phi_2=0$. The ``Random" configuration refers to 
unpolarized U+U collisions. For Au+Au collisions that include small oblate deformation, we do not find significant 
difference between specific orientations and only show the distribution for the unpolarized collisions.  

The distributions for Au+Au and Tip-Tip U+U collisions correspond to smaller values of maximal $|L-R|$ as compared 
to random U+U configuration. This is because, in these cases, non-zero values of $|L-R|$ arise 
purely from quantum fluctuations of the nucleon positions. The small difference in the probability distributions 
between these configurations is due to the larger number of nucleons in Uranium that broadens the probability 
distribution of $\l L-R\r$ itself. The Body-Tip configuration, as expected, produces the highest 
spectator neutron asymmetry. Thus as argued before and as shown in Fig.~\ref{fig.norm_dist}, in case of Random U+U 
collisions, events with highest values of $|L-R|$ are dominated by Body-Tip events. For the rest part of this paper, 
by U+U collisions we refer to ``Random U+U" collisions. 

We will now study the variation of ellipticity and CME like signal with $|L-R|$. In our calculation 
we define ellipticity ($\varepsilon_2$) as 
\begin{equation}
\varepsilon_2 e^{i2\Psi_2^{{PP}}} = \frac{\sum\limits_{p} r_p^2 e^{i 2\phi_p}}{\sum\limits_{p}r_p^2},  
\end{equation}
where the sum is performed over the positions of participant nucleons (assumed to be delta-functions) at positions 
($r_p,\phi_p$). Here $\Psi_2^{{PP}}$ denotes the second order participant plane~\cite{PhysRevC.81.054905,
Teaney:2010vd,Qin:2010pf,Qiu:2012uy}, a measure of the direction of the initial state spatial anisotropy. 
In order to serve as a proxy for the experimental event plane, that measures the final state momentum anisotropy, 
$\Psi_2^{{PP}}$ is rotated by $\pi/2$ as per convention~\cite{Qin:2010pf}. A similar plane can be defined using the 
positions of spectator neutrons ($r_s,\phi_s$) in the collision plane as 
\begin{equation}
\Psi_{2}^{SP} = \frac{1}{2} \tan^{-1} \left( \frac{\sum\limits_{s} r_s^2 \sin 2\phi_s}{\sum\limits_{s} r_s^2 
\cos 2\phi_s} \right).
\end{equation}
In experiment, the plane of spectator neutrons can be directly measured using ZDCs~\cite{Abelev:2009ad, gang_wang}. 

The other quantity of interest is the correlation of the magnetic field $B$ generated by the protons in 
the U nuclei with the reaction plane $\Psi_{RP}$  defined as 
\begin{equation}
\gamma^B=e^2B^2\cos\l2\l\Psi_B-\Psi_{RP}\r\r.
\end{equation}
Here $B$ is the magnitude and $\Psi_B$ is the angle of the resultant $B$-field vector at the point of 
observation due to all protons. This correlator has been argued to serve as a proxy for the CME signal 
$\gamma^{ab}$~\cite{Bloczynski:2012en,Bloczynski:2013mca}. This quantity is similar to Eq.\ref{eq.cosab} 
of $\gamma^{ab}$ with $\phi_a + \phi_b$ replaced by $2\Psi_B$.

In this work we estimate the values of $B$ and $\Psi_B$ at the time of collision ($t=0$) 
using the expression of Lienard-Wiechert potentials~\cite{Landau_book,Skokov:2009qp,Deng:2012pc}. 
The estimates shown here are at the central point of the participant zone defined by the average weighted 
positions of the participants. We compute $\varepsilon_2$ and $\gamma^B$ for different 
bins of $|L-R|$ in a given centrality class. The centrality selection in our case is done from the min-bias 
multiplicity distribution of U+U collisions.
  \begin{figure}[htb]
    \centering
    \scalebox{1}{
\includegraphics[width=0.45\textwidth]{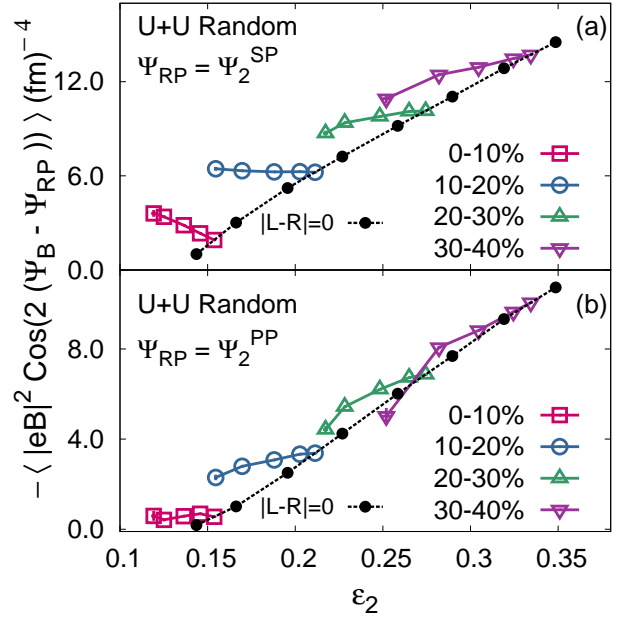}
    }
  \caption{(Color online) The correlator $\gamma^B$ at the centre of the participant zone vs $\varepsilon_2$ for 
  different centrality and spectator asymmetry bins. {\it{Top:}} $\gamma^B$ measured with respect to the 2nd 
  spectator plane. {\it{Bottom:}} $\gamma^B$ measured with respect to the 2nd participant plane. Bins with same 
  $\varepsilon_2$ is observed to have multiple $\gamma^B$ and vice-versa.
    }
  \label{fig.final}
    \end{figure}
\begin{figure*}[htb]
  \begin{center}
    \scalebox{1}
    {
   \hspace{-12pt} \includegraphics[width=1.1\textwidth]{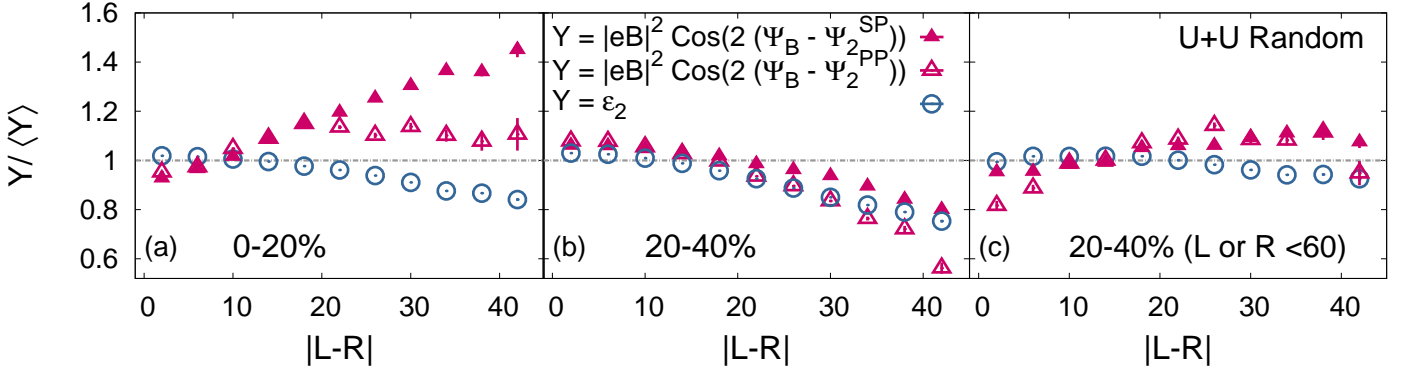}
    }
  \caption{(Color online) The initial state $\varepsilon_2$ and $\gamma^B$ 
  at the origin of the participant plane vs $|L-R|$ for different choices of $\psi_{RP}$, $\psi_{RP}=\psi^{PP}_2$ 
  and $\psi^{SP}_2$. The plots are for $\l0-20\r\%$ and $\l20-40\r\%$ centrality cases. Panel (c) again shows the 
  $\l20-40\r\%$ centrality class with additional cut on the maximum value of $L$ or $R$}
  \label{fig.0to20cent}
  \end{center}
  \end{figure*}

In Fig.~\ref{fig.final} we show the correlation between $\gamma^B$ and $\varepsilon_2$ 
for different bins of centrality and $|L-R|$. We have used $\Psi_{RP}=\Psi_{2}^{SP}$ (Fig.\ref{fig.final}.(a)) and 
$\Psi_{2}^{PP}$ (Fig.\ref{fig.final}.(b)). The centrality is varied in the range of $0-40\%$ over which we expect 
$\varepsilon_2$ to be a good proxy for $v_2$~\cite{Qiu:2011iv,Schenke:2013aza}. The $|L-R|=0$ bins for different 
centralities are shown by solid symbols. This indicates that $\gamma^B$ proportionally increases with $\varepsilon_2$ 
for different centrality bins. Thus by changing the centralities alone it is not possible to disentangle CME from that 
of the collective response to $\varepsilon_2$. An interesting behavior is seen by introducing the spectator asymmetry 
parameter which allows us to tune $\varepsilon_2$ in a given centrality class. The different points in a given centrality 
bin, shown by open symbols are obtained when $|L-R|$ is varied in the range of 0 to 45. In all the cases the different 
curves extend out from the $|L-R|=0$ curve. This indicates that a given value of $\varepsilon_2$ corresponds to multiple 
values of $\gamma^B$ and vice-versa. This is more prominent for the two most central bins $\l0-10\r\%$ and $\l10-20\r\%$ 
and much stronger when $\Psi_{2}^{SP}$ is used as $\Psi_{RP}$. 

We argue that this behavior can be the key to disentangle the effects of $\gamma^B$ that drives $\gamma^{ab}$ from initial 
$\varepsilon_2$ that drives $v_2$. A correlation plot of $\gamma^{ab}$ vs $v_2$ for different centralities for $|L-R|=0$,  
similar to Fig.\ref{fig.final} can be measured in the experiment. The next step would be to extract the same correlation 
by varying $|L-R|$ at a given centrality. If $\gamma^{ab}$ is only sensitive to the variation in $v_2$ the two correlation 
curves will overlap. On the other hand, if $\gamma^{ab}$ is driven by CME, one expects different slopes of the correlation 
curves as shown in Fig.\ref{fig.final}. In that case, a given value of $v_2$ will correspond to multiple values of 
$\gamma^{ab}$ and vice-versa. Such observation will lead to a clear disentanglement between the two effects and a strong 
support for CME.

We note from Fig.~\ref{fig.final} that there is a hint of anti-correlation between $\varepsilon_2$ and $\gamma^B$ for the 
central collisions. This is clearly visible in Fig.~\ref{fig.final} (a). We try to focus on this aspect more elaborately 
in Fig.~\ref{fig.0to20cent}. Here we have combined the two central bins $\l0-10\r\%$ and $\l10-20\r\%$ into a single bin 
of $\l0-20\r\%$ and the rest into another bin of $\l20-40\r\%$. In each case we plot the variation of $\varepsilon_2$ and 
$\gamma^B$ scaled by their mean values in a given centrality bin with respect to $|L-R|$. For both centralities, $\varepsilon_2$ 
monotonically decreases with increasing $|L-R|$ by about 20$\%$. Interestingly, as shown in Fig.\ref{fig.0to20cent} (a), 
for the $\l0-20\r\%$ centrality bin $\gamma^B$ grows with increasing $|L-R|$. The growth is stronger when $\Psi_2^{SP}$ 
is used for the estimation of event plane. In that case about $50\%$ increase is observed. When $\Psi_2^{PP}$ is used for 
the calculation, $\gamma^B$ initially increases by about $20\%$ and becomes nearly constant. Thus for the $\l0-20\r\%$ 
centrality bin, we find that by varying $|L-R|$ alone one can in principle observe anti-correlation between the effects of 
$B-$field that lead to CME-like signals and flow-like effects. However we find that for the $\l20-40\r\%$ centrality bin, 
both $\varepsilon_2$ and $\gamma^B$ go down with increasing $|L-R|$. This is shown in Fig.~\ref{fig.0to20cent} (b). This 
trend however drastically changes (as shown in Fig.~\ref{fig.0to20cent} (c)) by restricting the number of spectators on 
any one side ($L$ or $R$). We find that by applying a cut of $L$ or $R$ less than 60, once again we see a trend qualitatively 
similar to the $\l0-20\r\%$ bin. Such observation of anti-correlation between $\gamma^{ab}$ and $v_2$ in data will lead to a 
very strong case for CME.

The effects seen in Fig.~\ref{fig.final} and Fig.~\ref{fig.0to20cent} can be understood in the following way. As illustrated 
in Figs.~\ref{fig.scheme} and \ref{fig.norm_dist}, increasing $|L-R|$ in general increases the fraction of Body-Tip events 
thereby decreasing $\varepsilon_2$ in a multiplicity bin. This decrease of $\varepsilon_2$ with $|L-R|$ is a robust effect 
intrinsic to the prolate deformation of Uranium. However the strength of the projected $B-$field 
$e^2B^2\cos\l2\l\Psi_B-\Psi_{RP}\r\r$ have a more complicated dependence on $|L-R|$. For the central collisions, increasing 
$|L-R|$ also results in increasing the number of spectators. This is because the smaller $|L-R|$ bins in central events are 
dominated by full overlap collisions with very few spectators compared to central Body-Tip events. Thus, a relative increase 
of spectators is obtained by increasing $|L-R|$. However, in the $\l20-40\r\%$ bin increasing $|L-R|$ does not necessarily 
increase the number of spectators. In this centrality, the smaller $|L-R|$ bins are already dominated by non-central 
collisions which have large number of spectators. Introduction of an additional cut on the number of spectators on either 
side $L$ or $R$, once again leads to a relative increase of spectator and therefore an increase of $\gamma^B$ with $|L-R|$.  

In summary, we find that in U+U collisions the spectator neutron asymmetry $|L-R|$ gives us a direct access to the 
initial state geometry. It serves as a control parameter to trigger events with different 
values of initial anisotropy. This parameter can be experimentally measured by using a combination of ZDCs. We demonstrate 
spectator neutron asymmetry can be used to disentangle flow background from the signals of CME. We find that 
the correlation between projected $B-$field and ellipticity by varying centrality is different from the one obtained by 
varying spectator neutron asymmetry. The difference in the nature of these two correlations can be used to disentangle 
the signals of CME from flow. An experimental confirmation on the studies presented here will lead to a strong support 
for CME. 

This new method to constrain the initial state geometry could be a crucial input for various studies. The main advantage 
of spectator asymmetry is that it is not affected by the evolution of the system and the final state interactions. This 
will allow us to perform precision studies of the influence of the initial state geometry on the hydrodynamic response 
of the fireball~\cite{Teaney:2010vd,Bhalerao:2011bp}, jet suppression etc~\cite{Heinz:2004ir,Zhang:2012ie,Zhang:2012ha}.

{\it Acknowledgement:} We acknowledge our discussions with R. Haque, H. Masui, B.  Mohanty, P. Sorensen and G. Wang. 
We are grateful to B. Schenke for his comments on the manuscript. The DRONA and PRAFULLA clusters of Computer Division 
and the LHC grid computing centre at the Variable Energy Cyclotron Centre, supported by the Department of Atomic Energy,
Government of India are acknowledged. 
\bibliographystyle{apsrev4-1}
\bibliography{CME}

\end{document}